\documentclass[sigconf]{acmart}
\usepackage[binary-units,per-mode=symbol,per-symbol=p]{siunitx}
\usepackage[utf8]{inputenc}
\usepackage{adjustbox}
\usepackage[ruled,vlined]{algorithm2e}
\usepackage{algorithmic}
\usepackage{graphicx}
\usepackage{textcomp}
\usepackage{xcolor}
\usepackage{balance}
\usepackage{enumitem}
\usepackage{setspace}
\usepackage{wrapfig}
\usepackage{listings}
\usepackage{color}
\usepackage{caption}
\usepackage{subcaption}

\usepackage{multirow}
\usepackage{multicol}

\usepackage{tikz}
\usetikzlibrary{shapes.geometric, arrows}
\usetikzlibrary{fit}
\usetikzlibrary{calc}
\tikzstyle{arrow} = [thick,->,>=stealth]

\usepackage{amsmath}

\usepackage{mathtools}

\usepackage{ctable}
\newcommand{\etal}{\emph{et al.}\xspace}
\newcommand{\ie}{\emph{i.e.}, }
\newcommand{\eg}{\emph{e.g.}, }

\newcommand{\cf}{\emph{cf.}\xspace}
\newcommand{\HAS}{\emph{HTTP Adaptive Streaming }}
\newcommand{\DASH}{\emph{Dynamic Adaptive Streaming over HTTP }}
\newcommand{\HEVC}{\emph{High Efficiency Video Coding }}
\newcommand{\VVC}{\emph{Versatile Video Coding }}

\newcommand{\DCT}{\emph{Discrete Cosine Transform }}

\newcommand{\HLS}{\emph{HTTP Live Streaming }}

\newcommand{\caps}{\texttt{CAPS}\xspace}
\newcommand{\ztps}{\texttt{JALE}\xspace}

\newcommand{\EY}{$E_{\text{Y}}$}
\newcommand{\LY}{$L_{\text{Y}}$}
\newcommand{\h}{$h$}

\newcommand{\BDRP}{BDR\textsubscript{P}}
\newcommand{\BDRV}{BDR\textsubscript{V}}
\newcommand{\vJ}{$v_{\text{J}}$\xspace}

\copyrightyear{2024}
\acmYear{2024}
\setcopyright{rightsretained}
\acmConference[MHV '24]{Mile-High Video Conference}{February 11--14, 2024}{Denver, CO, USA}
\acmBooktitle{Mile-High Video Conference (MHV '24), February 11--14, 2024, Denver, CO, USA}
\acmDOI{10.1145/3638036.3640807}
\acmISBN{979-8-4007-0493-2/24/02}

\begin{document}

\title{Optimal Quality and Efficiency in Adaptive Live Streaming with JND-Aware Low latency Encoding}

\author{Vignesh V Menon}
\email{vignesh.menon@hhi.fraunhofer.de}
\orcid{0000-0003-1454-6146}
\affiliation{
  \institution{\small{Video Communication and Applications Dept}}
  \institution{Fraunhofer HHI}
  \city{Berlin}
  \country{Germany}
}

\author{Jingwen Zhu}
\email{jingwen.zhu@etu.univ-nantes.fr}
\affiliation{
  \institution{\small{Nantes Université}}
  \institution{École Centrale Nantes, CNRS, LS2N, UMR 6004}
  \city{Nantes}
  \country{France}
}

\author{Prajit T Rajendran}
\email{prajit.thazhurazhikath@cea.fr}
\orcid{0000-0002-8283-9891}
\affiliation{
  \institution{\small{CEA, List, F-91120 Palaiseau}}
  \institution{Université Paris-Saclay}
  \city{Paris}
  \country{France}
}

\author{Samira Afzal}
\email{samira.afzal@aau.at}
\affiliation{
  \institution{\small{Institute of Information Technology}}
  \institution{Alpen-Adria-Universität}
  \city{Klagenfurt}
  \country{Austria}
}

\author{Klaus Schoeffmann}
\email{klaus.schoeffmann@aau.at}
\affiliation{
  \institution{\small{Institute of Information Technology}}
  \institution{Alpen-Adria-Universität}
  \city{Klagenfurt}
  \country{Austria}
}

\author{Patrick Le Callet}
\email{patrick.lecallet@univ-nantes.fr}
\affiliation{
  \institution{\small{Nantes Université}}
  \institution{École Centrale Nantes, IUF, CNRS, LS2N, UMR 6004}
  \city{Nantes}
  \country{France}
}

\author{Christian Timmerer}
\email{christian.timmerer@aau.at}
\orcid{0000-0002-0031-5243}
\affiliation{
  \institution{\small{Institute of Information Technology}}
  \institution{Alpen-Adria-Universität}
  \city{Klagenfurt}
  \country{Austria}
}

\renewcommand{\shortauthors}{Vignesh V Menon~\etal}

\begin{abstract}
In HTTP adaptive live streaming applications, video segments are encoded at a fixed set of bitrate-resolution pairs known as \textit{bitrate ladder}. Live encoders use the fastest available encoding configuration, referred to as \textit{preset}, to ensure the minimum possible latency in video encoding. However, an optimized preset and optimized number of CPU threads for each encoding instance may result in \textit{(i)}~increased quality and \textit{(ii)}~efficient CPU utilization while encoding. For \textit{low latency} live encoders, the encoding speed is expected to be more than or equal to the video framerate. 
To this light, this paper introduces a \textbf{J}ust Noticeable Difference (JND)-\textbf{A}ware \textbf{L}ow latency \textbf{E}ncoding Scheme (\ztps), which uses random forest-based models to jointly determine the optimized encoder preset and thread count for each representation, based on video complexity features, the target encoding speed, the total number of available CPU threads, and the target encoder. 
Experimental results show that, on average, \ztps~yield a quality improvement of \SI{1.32}{\decibel} PSNR and 5.38~VMAF points with the same bitrate, compared to the fastest preset encoding of the HTTP Live Streaming (HLS) bitrate ladder using x265 HEVC open-source encoder with eight CPU threads used for each representation. These enhancements are achieved while maintaining the desired encoding speed. Furthermore, on average, \ztps results in an overall storage reduction of \SI{72.70}{\percent}, a reduction in the total number of CPU threads used by \SI{63.83}{\percent}, and a \SI{37.87}{\percent} reduction in the overall encoding time, considering a JND of six VMAF points.
\end{abstract}



\maketitle

\section{Introduction}
The Moving Picture Experts Group (MPEG) has developed a standard called \DASH (MPEG-DASH) \cite{DASH_IEEE} to meet the high demand for streaming high-quality video content over the Internet and overcome the associated challenges in \HAS (HAS)~\cite{has_ref}. The main idea behind HAS is to divide the video content into \textit{segments} and to encode each segment at various bitrates and resolutions, called \textit{representations}. These representations enable a continuous adaptation of the video delivery to the client's network conditions and device capabilities~\cite{DASH_Survey}. The increase in the computational complexity using codecs such as \HEVC~(HEVC)~\cite{HEVC} and \VVC~(VVC)~\cite{VVC}, and improvements in video characteristics such as resolution~\cite{jtps_ref}, framerate~\cite{cvfr_ref}, and bit-depth raises the need to develop a large-scale, highly efficient video encoding environment~\cite{netflix_paper}. This is crucial for DASH-based content provisioning as it requires encoding multiple representations of the same video content in an \emph{encoding server}.

\textit{\textbf{Motivation:}} 
Traditionally, a fixed bitrate ladder, \eg \HLS (HLS) bitrate ladder~\cite{HLS_ladder_ref}, is used in live streaming applications. Furthermore, for every representation, maintaining an encoding speed that is the same as or greater than the video framerate, regardless of the complexity of the video content, is a crucial goal for a low latency live encoder~\cite{pradeep_ref}. Although the output video's compression efficiency (in terms of the obtained perceptual quality and bitrate) is an essential metric for the encoder, maintaining the encoding speed takes precedence in live streaming scenarios. This is because a reduction in encoding speed may lead to the unacceptable outcome of dropped frames during transmission, eventually decreasing the quality of experience~\cite{pradeep_ref}. The encoding speed depends on video content complexity and parameters such as \textit{(i)} target resolution, \textit{(ii)} target bitrate, \textit{(iii)} number of CPU threads~\cite{xiao2015hevc}, and \textit{(iv)} encoder configuration~\cite{ds_paper_ref}.

\begin{figure}[t]
    \centering
    \includegraphics[width=0.432\textwidth]{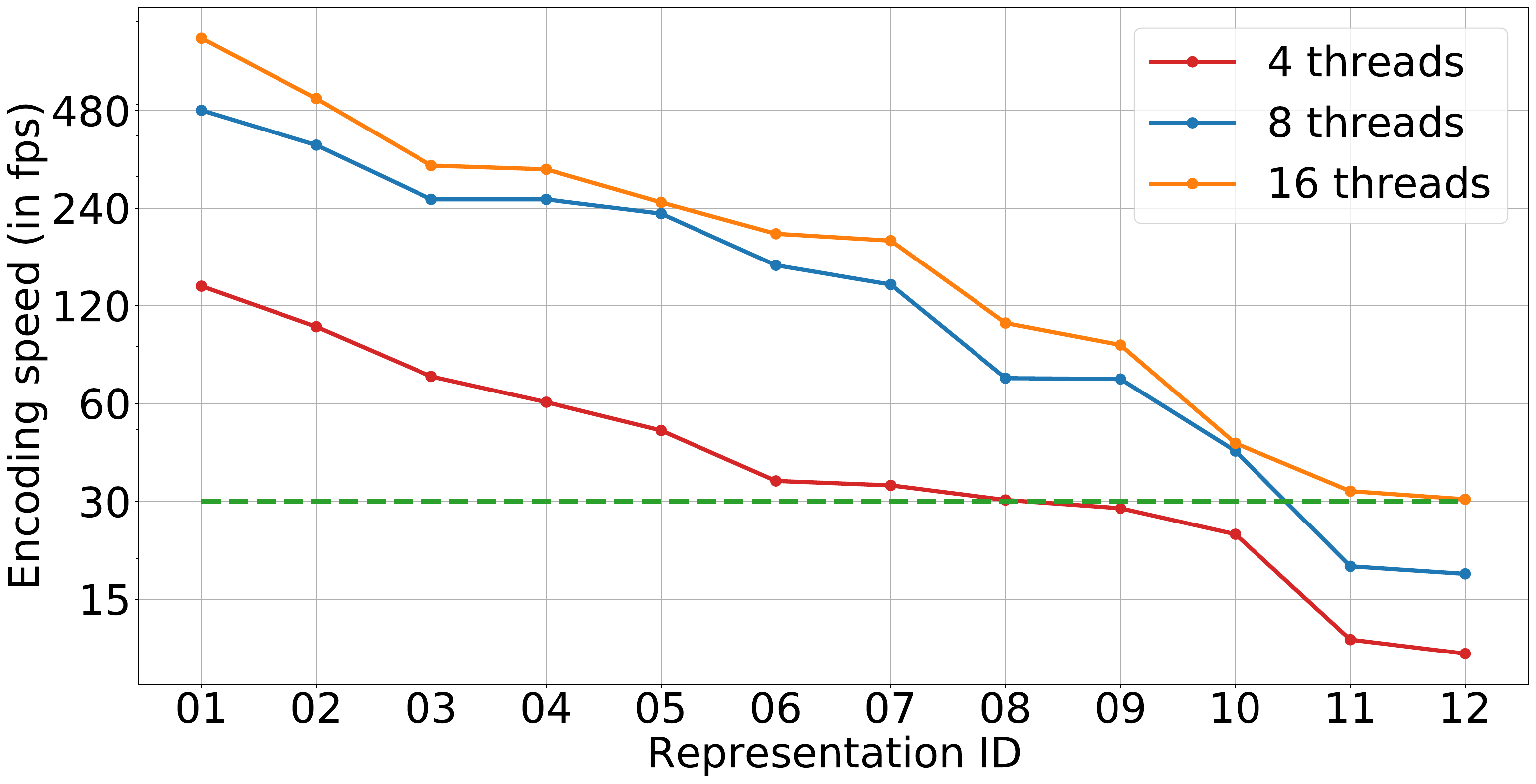}
    \caption{The encoding speed of each representation in HLS bitrate ladder~\cite{HLS_ladder_ref} for the \textit{Wood\_s000} sequence~\cite{VCD_ref} using \textit{ultrafast} preset of x265~\cite{x265_ref} using 4, 8, and 16 CPU threads for each representation.}
    \vspace{-0.97em}
    \label{fig:motive_eg}
\end{figure}

\textit{Optimized resource allocation in encoding servers:} In adaptive streaming, strategically allocating an optimized number of CPU threads for each video encoder instance at a cloud server is crucial. Tailoring CPU thread count to encoding resolution and target bitrate allows for precise resource allocation, enhancing the efficiency of the encoding process. The cloud server can significantly streamline the encoding process by dynamically adjusting CPU thread counts based on resolution and bitrate, accommodating diverse video qualities within an adaptive streaming environment. Figure~\ref{fig:motive_eg} shows the encoding time measurement of an entire HLS HEVC bitrate ladder encoding of the \textit{Wood\_s000} sequence~\cite{VCD_ref} using \textit{ultrafast} preset of x265~\cite{x265_ref} with 4, 8, and 16 CPU threads for each of the twelve encoding representations. Since the video (segment) is of 30\,fps, the target encoding speed is considered 30\,fps~\cite{pradeep_ref}. However, specific thread configurations cannot deliver certain representations' desired encoding speed of 30 fps. For example, the 11.6\,Mbps and 16.8\,Mbps representations do not achieve the target encoding speed (30\,fps) using 4 or 8 threads. This emphasizes the impact of CPU utilization and thread configuration on the encoding performance.

\begin{figure*}[t]
\centering
\includegraphics[width=0.881\linewidth]{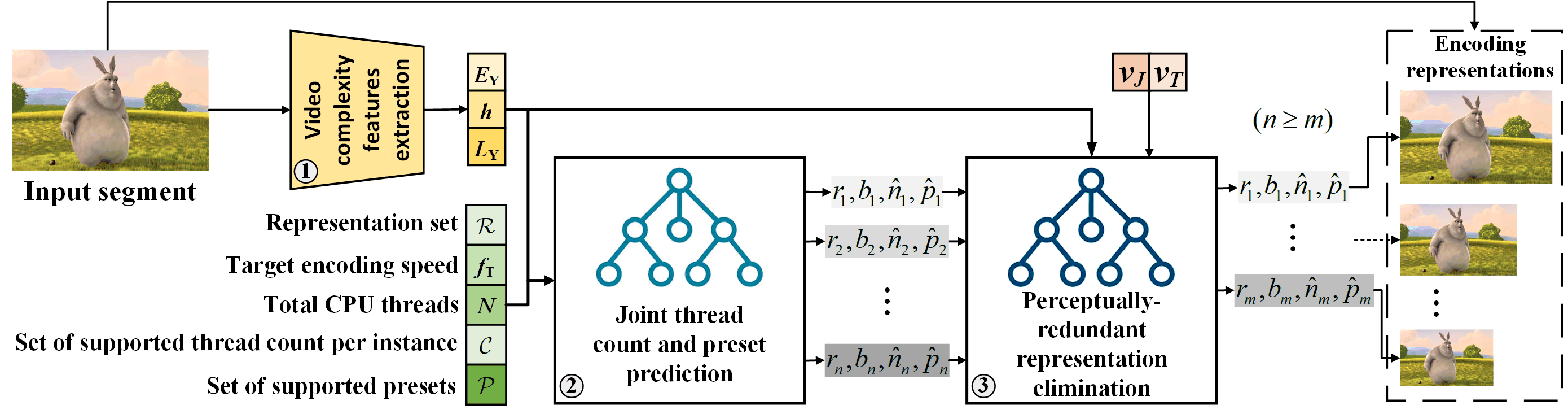}
\caption{Live encoding using \ztps envisioned in this paper.}
\label{fig:contribution}
\end{figure*}

\textit{Optimized encoding preset:} Traditional open-source encoders like x264~\cite{x264_ref}, x265~\cite{x265_ref}, and VVenC~\cite{vvenc_ref} have pre-defined sets of encoding parameters (termed as \textit{presets}), which present a trade-off between the encoding time and compression efficiency~\cite{Silveira2017Performance,cvfr_ref}. The preset for the fastest encoding (\textit{ultrafast} for x264 and x265) is used as the encoder preset for the entire live content, independent of the video content complexity~\cite{pradeep_ref}. Moreover, the streaming service provider arbitrarily chooses the number of CPU threads for each representation, irrespective of the content complexity. Though the conservative technique of fixing the preset and thread count for each encoding instance may achieve the intended result of a low latency encoding, the resulting encoding is sub-optimal, especially when the type of the content is dynamically changing, which is the typical use case for live streams~\cite{perf_ref}. Furthermore, when the content becomes easier to encode (\ie slow-moving videos or videos that have simpler textures are easy to encode as predicting the current frame from a previous frame is simpler, resulting in smaller residuals), the encoder would achieve a higher encoding speed than the target encoding speed. This, in turn, introduces unnecessary CPU idle time as it waits for the video feed. If the encoder preset is configured such that this higher encoding speed can be reduced while still being compatible with the expected live encoding speed, the quality of the encoded content achieved by the encoder can be improved. Subsequently, when the content becomes complex again, the encoder preset needs to be reconfigured to move back to the faster configuration that achieves live encoding speed~\cite{sota_ref1,cvfr_ref}. By employing efficient storage techniques and removing unnecessary representations, the energy consumption associated with storing and transmitting redundant data can be minimized~\cite{quortex2022mission}.

\textit{\textbf{Contributions:}} 
This paper proposes a Just Noticeable Difference (JND)-aware low latency encoding scheme (\ztps) that \textit{jointly determines the CPU thread count and encoder preset configuration for each bitrate representation dynamically, adaptive to the video content to achieve low latency encoding.}
Content-aware features, \ie \DCT (DCT)-energy-based low-complexity spatial and temporal features, are extracted to determine video segments' characteristics, which random forest-based models use to predict optimized thread count and encoder preset for each representation to maintain the target encoding speed.  
\ztps~achieve the desired target encoding speed while maximizing compression efficiency and minimizing the total CPU threads used. Furthermore, based on JND, \ztps~removes perceptual redundancy between representations in the bitrate ladder.

\textit{\textbf{Paper outline:}} 
The remainder of this paper is organized as follows. Section \ref{sec:ztps_framework} describes the proposed \ztps~encoding architecture. In Section \ref{sec:evaluation}, the performance of \ztps~is evaluated, and Section \ref{sec:conclusion_future_dir} concludes the paper.

\section{\ztps~architecture}
\label{sec:ztps_framework}
The architecture of \ztps~for streaming applications is presented in Figure~\ref{fig:contribution}, according to which the number of threads and encoder preset for every segment in each representation of the bitrate ladder is predicted using spatiotemporal features of the input video segment, the target video encoding speed ($s_{\text{T}}$), the set of pre-defined supported thread count per instance ($\mathcal{C}$), and the set of pre-defined encoder presets ($\mathcal{P}$). The encoding process is carried out with the predicted encoder preset and the number of threads for each video segment. \ztps is classified into three steps:
\begin{enumerate}
    \item video complexity feature extraction,
    \item joint thread count and preset prediction,
    \item perceptually-redundant representation elimination.
\end{enumerate}

\subsection{Video complexity feature extraction}
\label{sec:features}
Predictive models can comprehensively understand the content complexity and characteristics by extracting relevant spatiotemporal features, such as motion vectors, texture patterns, and frame-to-frame differences~\cite{vvenc_qp_pred}. In this paper, three DCT-energy-based features~\cite{vca_ref}, the average luma texture energy~(\EY), the average gradient of the luma texture energy~(\h), and the average luminescence~(\LY), for each segment are extracted using open-source Video Complexity Analyzer (VCA)~\cite{vca_ref, jtps_ref}.

\subsection{Joint thread count and preset prediction}
\label{sec:preset_pred}
Selection of the optimized thread count-preset pair for each segment per representation based on the video content complexity is decomposed into two parts:
\begin{enumerate}
    \item train models to predict the encoding speed for each thread count-preset pair,
    \item develop a function to obtain the optimized thread count- preset pair for each representation.
\end{enumerate}  

The encoding speed of the $t^{th}$ representation of the input video segment ($s_{t}$) is modeled as a  function of the video content complexity features, the target representation (resolution $r_{t}$ and bitrate $b_{t}$)~\cite{cvfr_ref}, the number of threads $n_{t}$, and the encoder preset $p_{t}$, as shown in the equation:
\begin{equation}
s_{t} = f_{\text{S}}(E_{\text{Y}}, h, L_{\text{Y}}, r_{t}, b_{t}, n_{t}, p_{t}) 
\end{equation}
We use random forest models~\cite{rf_ref} to predict the encoding speed for each thread count-preset pair. $(P_{\text{T}}\times C_{\text{T}})$ models are trained, where $P_{\text{T}}$ and $C_{\text{T}}$ represent the number of encoding presets and the number of supported thread count per instance, respectively.

The optimized thread count-preset prediction function has a look-up table of $(\hat{n}_{t}, \hat{p}_{t})$ pairs. The supported encoder presets are chosen based on the target encoder and the preference of the streaming service provider. For example, presets ranging from \textit{ultrafast} to \textit{veryslow} can be chosen for x264 and x265 encoders. The set of possible thread counts ($\mathcal{C}$) for every encoder instance is input by the streaming service provider based on the encoding server architecture. The priority of $(\hat{n}_{t}, \hat{p}_{t})$ pairs is decided based on the following constraints:
\begin{enumerate}
    \item the achieved encoding speed $\hat{s}_{t}$ of the $t^{th}$ representation must be greater than or equal to the target encoding speed $s_{\text{T}}$, \ie $\hat{s}_{t} \geq s_{\text{T}}$.
    \item total number of CPU threads used for each representation is minimized.
\end{enumerate}

\tikzstyle{process_ms} = [rectangle, minimum width=4.2cm, minimum height=0.7cm, text centered, text width=4.1cm, draw=black, fill=blue!10]
\tikzstyle{ellipse_ms} = [rectangle, minimum width=2.4cm, minimum height=0.7cm, text centered, text width=2.3cm, draw=black, fill=red!10]
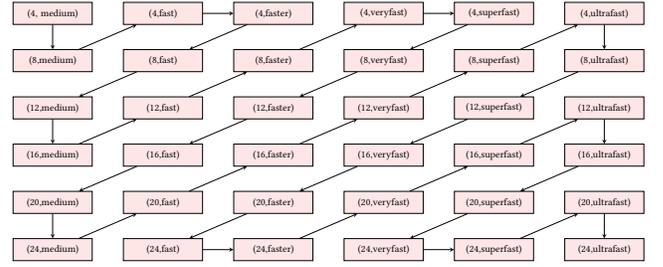
\begin{figure}[t]
\begin{center}
\resizebox{0.998\linewidth}{!}{
\begin{tikzpicture}[node distance=1.0cm]
\node (i1) [ellipse_ms] {(4, medium)};
\node (i2) [ellipse_ms, below of=i1, yshift = -0.5cm] {(8,medium)};
\node (i3) [ellipse_ms, below of=i2, yshift = -0.5cm] {(12,medium)};
\node (i4) [ellipse_ms, below of=i3, yshift = -0.5cm] {(16,medium)};
\node (i5) [ellipse_ms, below of=i4, yshift = -0.5cm] {(20,medium)};
\node (i6) [ellipse_ms, below of=i5, yshift = -0.5cm] {(24,medium)};

\node (j1) [ellipse_ms, right of=i1, xshift = 2.5cm] {(4,fast)};
\node (j2) [ellipse_ms, below of=j1, yshift = -0.5cm] {(8,fast)};
\node (j3) [ellipse_ms, below of=j2, yshift = -0.5cm] {(12,fast)};
\node (j4) [ellipse_ms, below of=j3, yshift = -0.5cm] {(16,fast)};
\node (j5) [ellipse_ms, below of=j4, yshift = -0.5cm] {(20,fast)};
\node (j6) [ellipse_ms, below of=j5, yshift = -0.5cm] {(24,fast)};

\node (k1) [ellipse_ms, right of=j1, xshift = 2.5cm] {(4,faster)};
\node (k2) [ellipse_ms, below of=k1, yshift = -0.5cm] {(8,faster)};
\node (k3) [ellipse_ms, below of=k2, yshift = -0.5cm] {(12,faster)};
\node (k4) [ellipse_ms, below of=k3, yshift = -0.5cm] {(16,faster)};
\node (k5) [ellipse_ms, below of=k4, yshift = -0.5cm] {(20,faster)};
\node (k6) [ellipse_ms, below of=k5, yshift = -0.5cm] {(24,faster)};

\node (l1) [ellipse_ms, right of=k1, xshift = 2.5cm] {(4,veryfast)};
\node (l2) [ellipse_ms, below of=l1, yshift = -0.5cm] {(8,veryfast)};
\node (l3) [ellipse_ms, below of=l2, yshift = -0.5cm] {(12,veryfast)};
\node (l4) [ellipse_ms, below of=l3, yshift = -0.5cm] {(16,veryfast)};
\node (l5) [ellipse_ms, below of=l4, yshift = -0.5cm] {(20,veryfast)};
\node (l6) [ellipse_ms, below of=l5, yshift = -0.5cm] {(24,veryfast)};

\node (m1) [ellipse_ms, right of=l1, xshift = 2.5cm] {(4,superfast)};
\node (m2) [ellipse_ms, below of=m1, yshift = -0.5cm] {(8,superfast)};
\node (m3) [ellipse_ms, below of=m2, yshift = -0.5cm] {(12,superfast)};
\node (m4) [ellipse_ms, below of=m3, yshift = -0.5cm] {(16,superfast)};
\node (m5) [ellipse_ms, below of=m4, yshift = -0.5cm] {(20,superfast)};
\node (m6) [ellipse_ms, below of=m5, yshift = -0.5cm] {(24,superfast)};

\node (n1) [ellipse_ms, right of=m1, xshift = 2.5cm] {(4,ultrafast)};
\node (n2) [ellipse_ms, below of=n1, yshift = -0.5cm] {(8,ultrafast)};
\node (n3) [ellipse_ms, below of=n2, yshift = -0.5cm] {(12,ultrafast)};
\node (n4) [ellipse_ms, below of=n3, yshift = -0.5cm] {(16,ultrafast)};
\node (n5) [ellipse_ms, below of=n4, yshift = -0.5cm] {(20,ultrafast)};
\node (n6) [ellipse_ms, below of=n5, yshift = -0.5cm] {(24,ultrafast)};

\draw [arrow] (i1) -- (i2);
\draw [arrow] (i2) -- (j1);
\draw [arrow] (j1) -- (k1);
\draw [arrow] (k1) -- (j2);
\draw [arrow] (j2) -- (i3);
\draw [arrow] (i3) -- (i4);
\draw [arrow] (i4) -- (j3);
\draw [arrow] (j3) -- (k2);
\draw [arrow] (k2) -- (l1);
\draw [arrow] (l1) -- (m1);
\draw [arrow] (m1) -- (l2);
\draw [arrow] (l2) -- (k3);
\draw [arrow] (k3) -- (j4);
\draw [arrow] (j4) -- (i5);
\draw [arrow] (i5) -- (i6);
\draw [arrow] (i6) -- (j5);
\draw [arrow] (j5) -- (k4);
\draw [arrow] (k4) -- (l3);
\draw [arrow] (l3) -- (m2);
\draw [arrow] (m2) -- (n1);
\draw [arrow] (n1) -- (n2);
\draw [arrow] (n2) -- (m3);
\draw [arrow] (m3) -- (l4);
\draw [arrow] (l4) -- (k5);
\draw [arrow] (k5) -- (j6);
\draw [arrow] (j6) -- (k6);
\draw [arrow] (k6) -- (l5);
\draw [arrow] (l5) -- (m4);
\draw [arrow] (m4) -- (n3);
\draw [arrow] (n3) -- (n4);
\draw [arrow] (n4) -- (m5);
\draw [arrow] (m5) -- (l6);
\draw [arrow] (l6) -- (m6);
\draw [arrow] (m6) -- (n5);
\draw [arrow] (n5) -- (n6);
\end{tikzpicture}
}
\end{center}
\vspace{-0.5em}
\caption{$(\hat{n}, \hat{p})$ look-up table used in the experimental validation of this paper.}
\label{fig:lut}
\end{figure}

An example look-up for x265 encoder is shown in Figure~\ref{fig:lut}, where $\mathcal{C}:\{4,8,12,16,20,24\}$, and $\mathcal{P}:$\{\emph{medium}, \emph{fast}, \emph{faster}, \emph{veryfast}, \emph{superfast}, \emph{ultrafast}\}. When the look-up table is scanned in the priority order, if the $(\hat{n}_{t}, \hat{p}_{t})$ pair yields an encoding speed higher than $f_{T}$, it is chosen as the optimized thread count-preset pair for the $t^{th}$ representation. 

\subsection{Perceptually-redundant representation elimination}
\label{sec:rep_elim}
\ztps~uses the JND-based representation elimination algorithm proposed in our previous work~\cite{cvfr_ref}. However, it is described in this paper (\cf Algorithm~\ref{algo:res_eliminate}) to make it self-contained. 
\begin{algorithm}[t]
\caption{JND-based representation elimination.}
\small
\textbf{Input:}\\
\quad $q$~: number of representations in $\mathcal{R}$\\
\quad $\mathcal{R}=\bigcup_{t=1}^q\left\{\left(r_{t},b_{t},\hat{n}_{t},\hat{p}_{t}\right)\right\}$:  \begin{flushright}representations with predicted thread count and preset \end{flushright}
\quad $\hat{v}_{t}~; 1 \leq t \leq q$: predicted VMAF\\
\quad $v_{\text{T}}$~: maximum VMAF threshold \\
\quad $v_{\text{J}}$~: average target JND \\
\textbf{Output:}  $\hat{\mathcal{R}}=\left(r,b, \hat{n}, \hat{p}\right)$: set of  encoding configurations\\
    $\hat{\mathcal{R}} \gets \left\{\left(r_{1},b_{1}, \hat{n}_{1}, \hat{p}_{1}\right)\right\}$  \\
    $u \gets 1$\\
    \If{$\hat{v}_{1} \geq v_{\text{T}}$}{
        \Return $\hat{\mathcal{R}}$
    }
        $t \gets 2$\\ 
        \While{$t \leq q$} {
            \If{$\hat{v}_{t} - \hat{v}_{u} \geq$ \vJ} {
                $\hat{\mathcal{R}} \gets \hat{\mathcal{R}}\cup \left\{\left(r_{t},b_{t}, \hat{n}_{t}, \hat{p}_{t}\right)\right\}$\\
                $u \gets t$  \label{alg:add_R_hat} \\
                \If{$\hat{v}_{t} \geq v_{\text{T}}$}{
                    \Return $\hat{\mathcal{R}}$
                }
            }
            $t \gets t + 1$
        }      
    \Return $\hat{\mathcal{R}}$
\label{algo:res_eliminate}
\end{algorithm}
The perceptual quality of the $t^{th}$ representation ($v_{t}$) is modeled as a function of the video content complexity features, the target representation (resolution $r_{t}$ and bitrate $b_{t}$), and the encoder preset $p_{t}$, as shown in the equation~\cite{cvfr_ref,mcbe_ref}: 
\begin{equation}
v_{t} = f_{\text{V}}(E_{\text{Y}}, h, L_{\text{Y}}, r_{t}, b_{t}, p_{t}) 
\end{equation}
Random forest models~\cite{rf_ref} are trained to predict the perceptual quality of each representation. 

\begin{table*}[t]
\caption{Experimental parameters of \ztps~used in this paper.}
\centering
\resizebox{\linewidth}{!}{
\begin{tabular}{l|c|c|c|c|c|c|c|c|c|c|c|c|c|c}
\specialrule{.12em}{.05em}{.05em}
\specialrule{.12em}{.05em}{.05em}
\multicolumn{2}{c|}{\emph{Parameter}} &  \emph{Symbol} & \multicolumn{12}{c}{\emph{Values}}\\
\specialrule{.12em}{.05em}{.05em}
\specialrule{.12em}{.05em}{.05em}
\multirow{2}{*}{\emph{Set of representations}} & \emph{Resolution height [pixels]} & \multirow{2}{*}{$\mathcal{R}$} & 360 & 432 & 540 & 540 & 540 & 720 & 720 & 1080 & 1080 & 1440 & 2160 & 2160 \\
& \emph{Bitrate [Mbps]} &  & 0.145 & 0.300 & 0.600 & 0.900 & 1.600 & 2.400 & 3.400 & 4.500 & 5.800 & 8.100 & 11.600 & 16.800 \\
\hline
\multicolumn{2}{c|}{\emph{Set of presets [x265]}} & $\mathcal{P}$ & \multicolumn{12}{c}{0 (\texttt{ultrafast}) -- 5 (\texttt{medium})}\\
\hline
\multicolumn{2}{c|}{\emph{Set of supported thread counts}} & $\mathcal{C}$ & \multicolumn{2}{c}{4} & \multicolumn{2}{|c}{8} & \multicolumn{2}{|c}{12} & \multicolumn{2}{|c}{16} & \multicolumn{2}{|c}{20} & \multicolumn{2}{|c}{24} \\
\hline
\multicolumn{2}{c|}{\emph{Total CPU threads}} & $N$ & \multicolumn{12}{c}{96}\\
\hline
\multicolumn{2}{c|}{\emph{Encoding speed threshold [fps]}} & $s_{\text{T}}$ & \multicolumn{12}{c}{30}\\
\hline
\multicolumn{2}{c|}{\emph{Average target JND}} & \vJ & \multicolumn{4}{c}{2} & \multicolumn{4}{|c}{4} & \multicolumn{4}{|c}{6}\\
\hline
\multicolumn{2}{c|}{\emph{Maximum VMAF threshold}} & $v_{\text{T}}$ & \multicolumn{4}{c}{98} & \multicolumn{4}{|c}{96} & \multicolumn{4}{|c}{94}\\
\specialrule{.12em}{.05em}{.05em}
\specialrule{.12em}{.05em}{.05em}
\end{tabular}
}
\label{tab:exp_par}
\end{table*}

This paper uses VMAF as the perceptual quality metric for each representation, while other quality metrics can be envisioned and are subject to future work. In practice, it is often observed that the VMAF scores of different representations are very similar, which introduces perceptual redundancy in the bitrate ladder.
To address this issue, this paper leverages the concept of the JND threshold, which represents the minimum threshold at which the human eye can perceive differences in quality~\cite{lin2015experimental, wang2016mcl, wang2017videoset}. The paper aims to eliminate perceptually redundant representations by utilizing the JND threshold. While~\cite{zhu2022framework, jasla_ref} have explored VMAF-based JND thresholds, their complexity is unsuitable for live-streaming applications. Therefore, this paper adopts a fixed JND threshold~\cite{kah_ref,jnd_streaming} denoted as $v_{\text{J}}$, an input from the streaming service provider. 
If the difference in the predicted VMAF of two representations is smaller than $v_{\text{J}}$, the representation with a higher bitrate will be eliminated.
If the predicted VMAF of a representation is larger than $v_{\text{T}}$, ($v_{\text{T}} = 100 - v_{\text{J}}$), \ie the threshold above which the representation is deemed perceptually lossless, the corresponding representation is eliminated from the bitrate ladder. In this manner, overall storage consumption and encoding energy of representations of an input video segment is reduced.

\begin{figure*}[t]
\centering
\begin{subfigure}{0.325\textwidth}
    \centering
    \includegraphics[width=\textwidth]{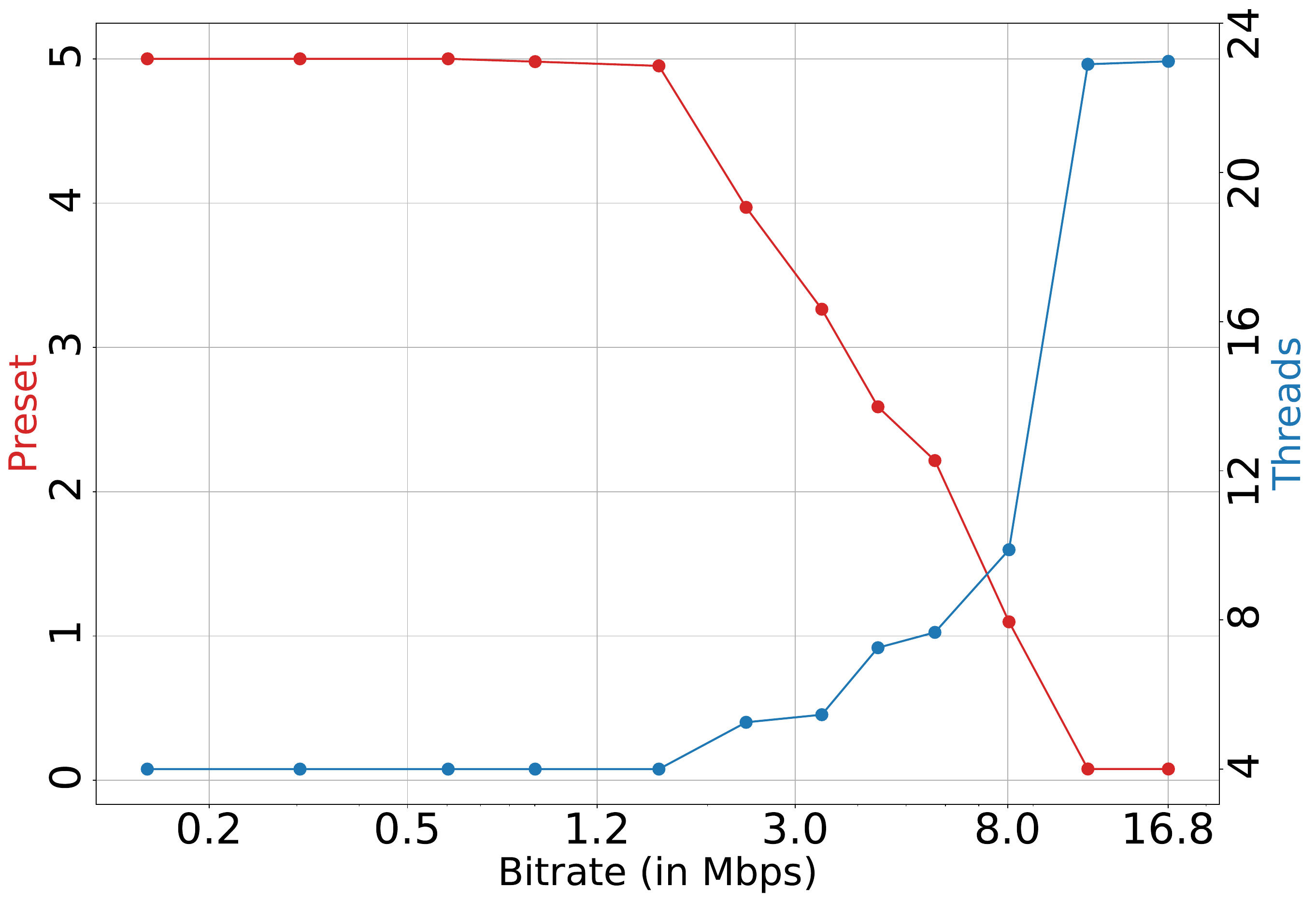}
    \caption{average thread count-preset}
    \label{fig:avg_preset}
\end{subfigure}
\hfill
\begin{subfigure}{0.31\textwidth}
    \centering
    \includegraphics[width=\textwidth]{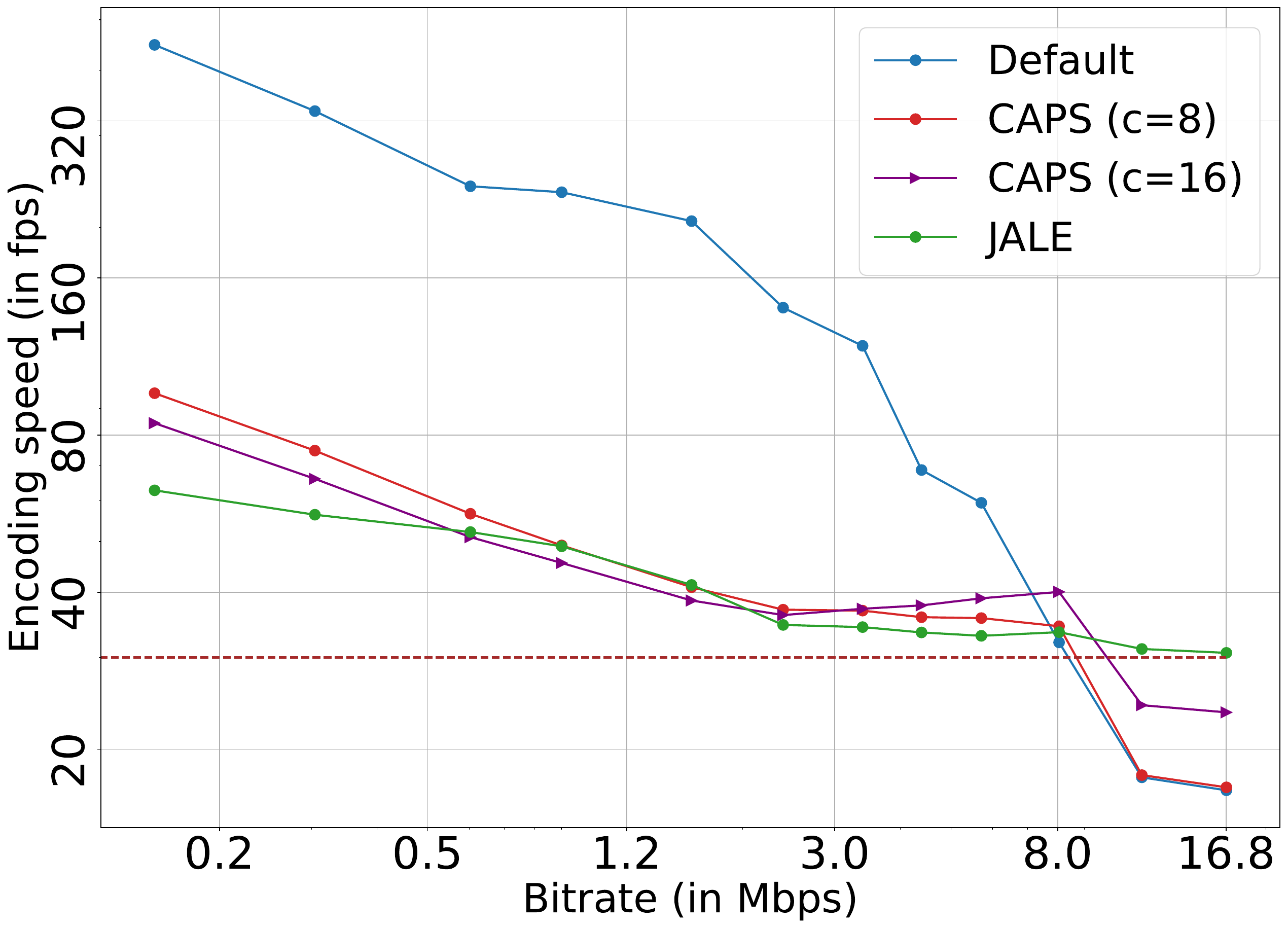}
    \caption{average encoding speed}
        \label{fig:time_rep}
\end{subfigure}
\hfill
\begin{subfigure}{0.31\textwidth}
    \centering
    \includegraphics[width=\textwidth]{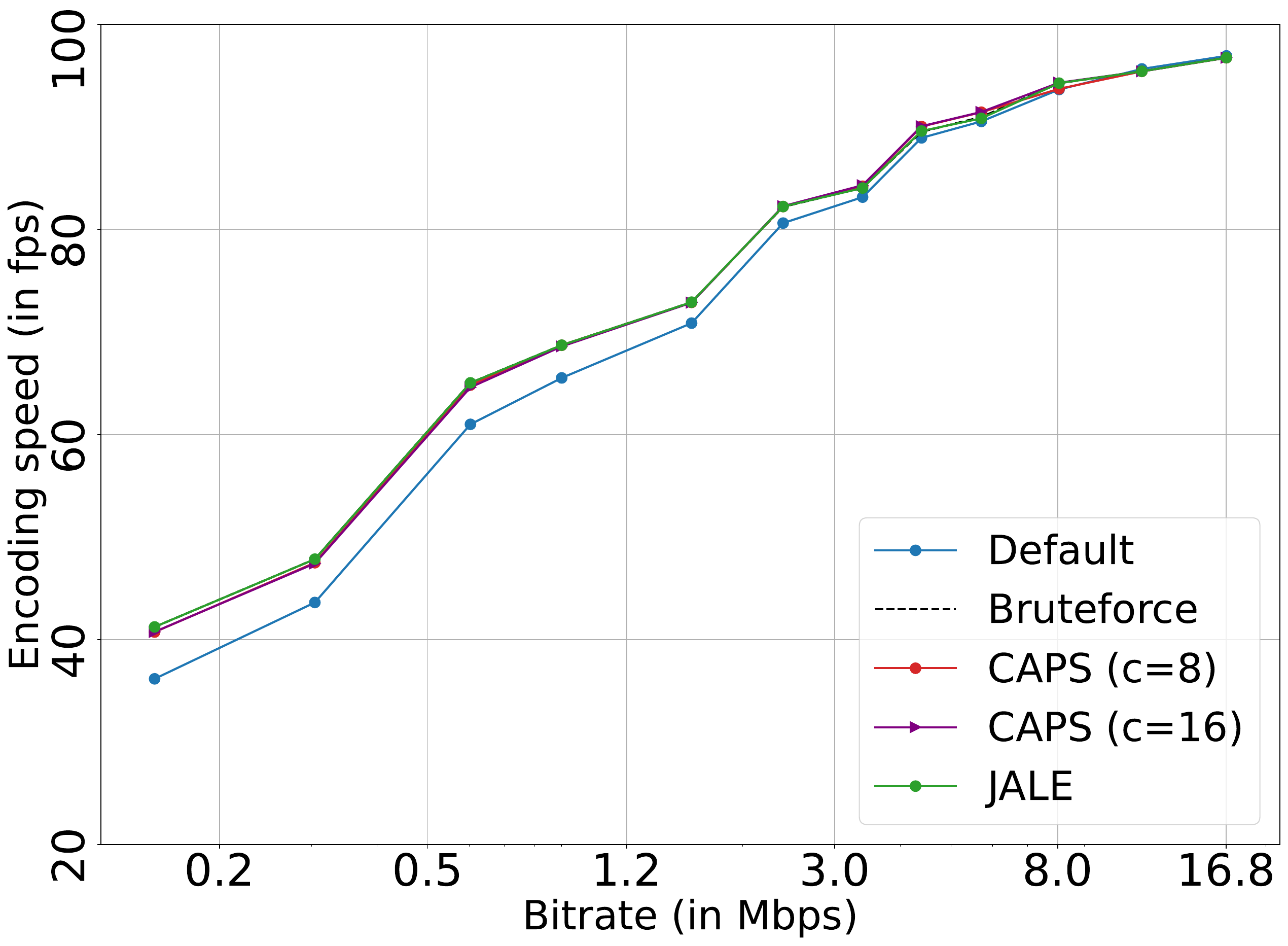}
    \caption{average VMAF}
    \label{fig:vmaf_rep}    
\end{subfigure}
\vspace{-0.3em}
\caption{Results for each representation in \ztps. JND-based representation elimination is not considered in these plots.}
\label{fig:main_plot}
\end{figure*}

\section{Evaluation}
\label{sec:evaluation}
This section introduces the test methodology used in this paper and presents the experimental results.

\subsection{Test Methodology}
\label{sec:test_methodology}
We use four hundred sequences (\SI{80}{\percent} of the sequences) from the video complexity dataset~\cite{VCD_ref} as the training dataset and the remaining (\SI{20}{\percent}) as the test dataset. We encode the sequences at 30\,fps using x265 v3.5~\cite{x265_ref} with multi-threading and x86 SIMD~\cite{x86_simd_ref} optimizations. The experimental parameters used in this paper are listed in Table~\ref{tab:exp_par}. We achieve CBR encoding for a target bitrate of $b_{t}$ (in Mbps) by setting the \textit{bitrate} and \textit{vbv-maxrate} option of x265 as $b_{t}$, and enabling \textit{strict-cbr} mode. 
We run all experiments on a dual-processor server with Intel Xeon Gold 5218R (80 cores, frequency at 2.10 GHz). We consider the bitrate ladder from HLS authoring specification for Apple devices~\cite{HLS_ladder_ref}. We consider $v_{\text{J}}$ as two~\cite{kah_ref}, four, and six~\cite{jnd_streaming} based on current industry practices. 

\begin{algorithm}[t]
\textbf{Inputs:}\\
\quad $\mathcal{Q}$: set of supported resolutions \\
\quad $\mathcal{B}$: set of target bitrates\\
\quad $\mathcal{C}$: set of thread count for each encoding instance\\
\For{each video segment}{
    Determine \{\EY, \h, \LY \} \\
    \For{each $r \in \mathcal{Q}$}{
        \For{each $b \in \mathcal{B}$}{
            \For{each $c \in \mathcal{C}$}{
            Encode segment with CBR $b$ \;
            Record \EY, \h, \LY, $r$, $b$, achieved bitrate $b'$, VMAF $v$, and PSNR $s$ \;
            }
        }
    }
}
\caption{Dataset generation.}
\label{algo:dataset_gen}
\end{algorithm}

\textit{\textbf{Prediction models:}} 
To ensure the robustness and generalization of the prediction models, we perform a five-fold cross-validation scheme for video sequences and average the results. The scheme also ensures that the test and training segments are split. We perform the hyperparameter tuning on the random forest prediction models on the \texttt{ultrafast} preset to balance the size and prediction accuracy of the models. The selected hyperparameters~\cite{scikit-learn} are \texttt{min\_samples\_leaf}= 1, \texttt{min\_samples\_split}= 2, \texttt{n\_estimators}= 100, \texttt{max\_depth}=14.

\textit{\textbf{Benchmark schemes:}} 
We compare \ztps~with the following encoding schemes:
\begin{enumerate}
    \item \emph{Default}: \textit{ultrafast} preset with eight threads for each encoding instance~\cite{HLS_ladder_ref}.
    \item \emph{Bruteforce}: optimized thread count-preset pair with and without JND-based representation elimination when the models are fully accurate. This is accomplished by bruteforce encoding using all thread count-preset pairs and selecting the optimized pair~\cite{netflix_paper}. Hence, it is suitable only for video-on-demand applications.
    \item \caps~\cite{caps_ref} determines the optimized preset for each representation for a target encoding speed of 30\,fps. We evaluate \caps where $c$=4, 8, and 16, respectively. 
\end{enumerate}

\textit{\textbf{Performance metrics:}} 
We compare \ztps~with the benchmark schemes using \BDRP{} and \BDRV~\cite{DCC_BJDelta}, which refers to the average increase in bitrate of the representations compared to the reference bitrate ladder encoding scheme to maintain the same PSNR and VMAF. A negative \mbox{BDR} suggests a boost in the coding efficiency of the considered encoding scheme compared to the reference bitrate ladder encoding scheme. Furthermore, we calculate \mbox{BD-PSNR} and \mbox{BD-VMAF}, which refer to the average increase in PSNR and VMAF at the same bitrate compared with the reference bitrate ladder encoding scheme. Positive \mbox{BD-PSNR} and \mbox{BD-VMAF} denote an increase in the coding efficiency of the considered encoding scheme compared to the reference bitrate ladder encoding. Relative storage space difference between the considered encoding scheme $b_{\text{opt}}$ and the reference encoding scheme $b_{\text{ref}}$ to store all bitrate ladder representations is evaluated as: 
    \begin{equation}
        \Delta S = \frac{\sum b_{\text{opt}}}{\sum b_{\text{ref}}} - 1.
    \end{equation}
We further determine the relative difference in the CPU thread count ($\Delta N$) needed for all bitrate ladder representations of the considered encoding schemes. We measure encoding energy consumption using the Running Average Power Limit (RAPL) interface and the \texttt{CodeCarbon} tool~\cite{codecarbon_ref}.

\subsection{Experimental Results}
\label{sec:exp_results}
\textit{\textbf{Predictions:}} 
Figure~\ref{fig:avg_preset} shows the average encoder preset and the number of CPU threads chosen for each representation of the bitrate ladder across the test dataset. On average, the 0.145\,Mbps representation chooses \textit{medium} preset ($p = 5$) and 4 threads, while 11.6\,Mbps and 16.8\,Mbps representations choose \textit{ultrafast} preset ($p = 0$) and 24 threads. This is because faster encoding presets and more computational resources are needed to encode high-bitrate representations, such that the encoding speed is above the threshold. 
Figure~\ref{fig:time_rep} shows the average encoding speed for every representation in the bitrate ladder using the \emph{default} encoding and \ztps. It is observed that, on average, the \ztps encodings have speeds above the lower bound of $s_{\text{T}}=30$\,fps. It is also observed that \emph{default} encoding has a high encoding speed for lower bitrate representations. This scenario occurs when the encoding tasks are relatively less demanding regarding CPU resources, allowing the CPU to remain substantially under-utilized while executing the encoding operations during a live feed. Furthermore, it could not achieve the minimum target encoding speed at higher bitrates (11.6\,Mbps and 16.8\,Mbps). However, \ztps controls the encoding speed to be greater than $s_{\text{T}}$ but not significantly higher than the \emph{default} encoding. This ensures higher CPU utilization when the encodings are carried out concurrently during a live feed.

\begin{figure*}[t]
\centering
\begin{subfigure}{0.244\textwidth}
    \centering
    \includegraphics[width=\textwidth]{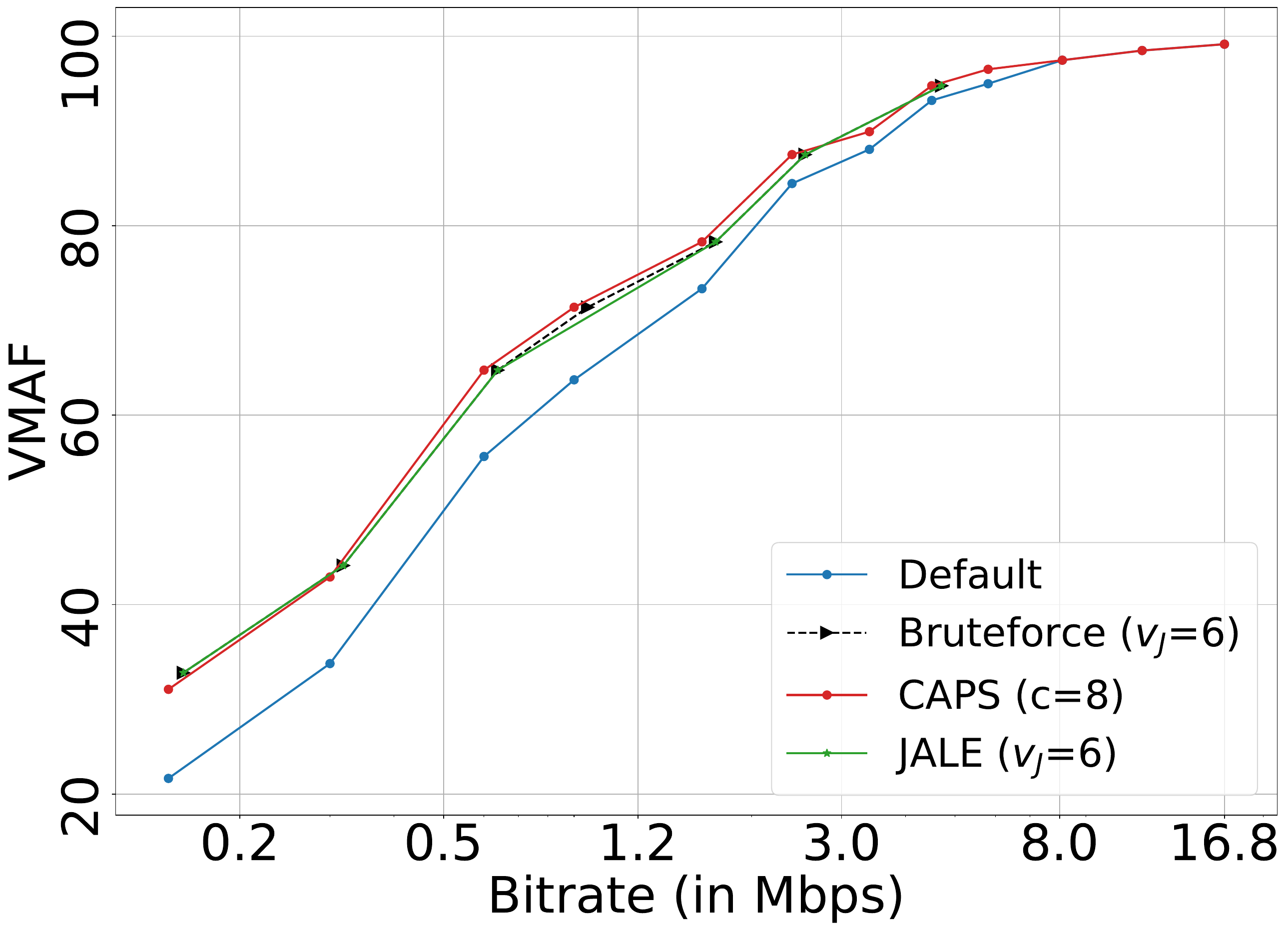}
    \includegraphics[width=\textwidth]{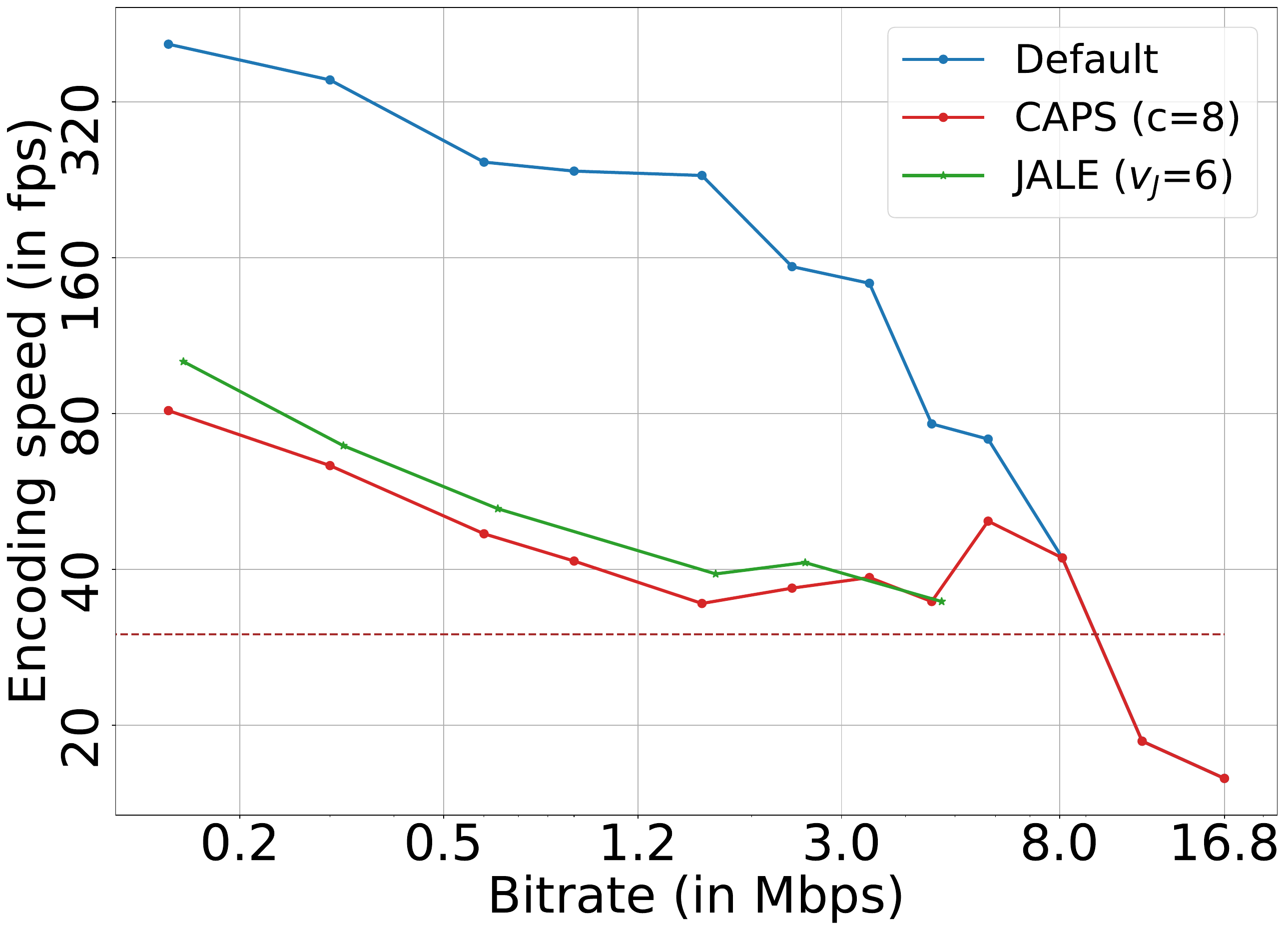}
    \caption{\textit{Basketball\_s000}}
\end{subfigure}
\hfill
\begin{subfigure}{0.244\textwidth}
    \centering
    \includegraphics[width=\textwidth]{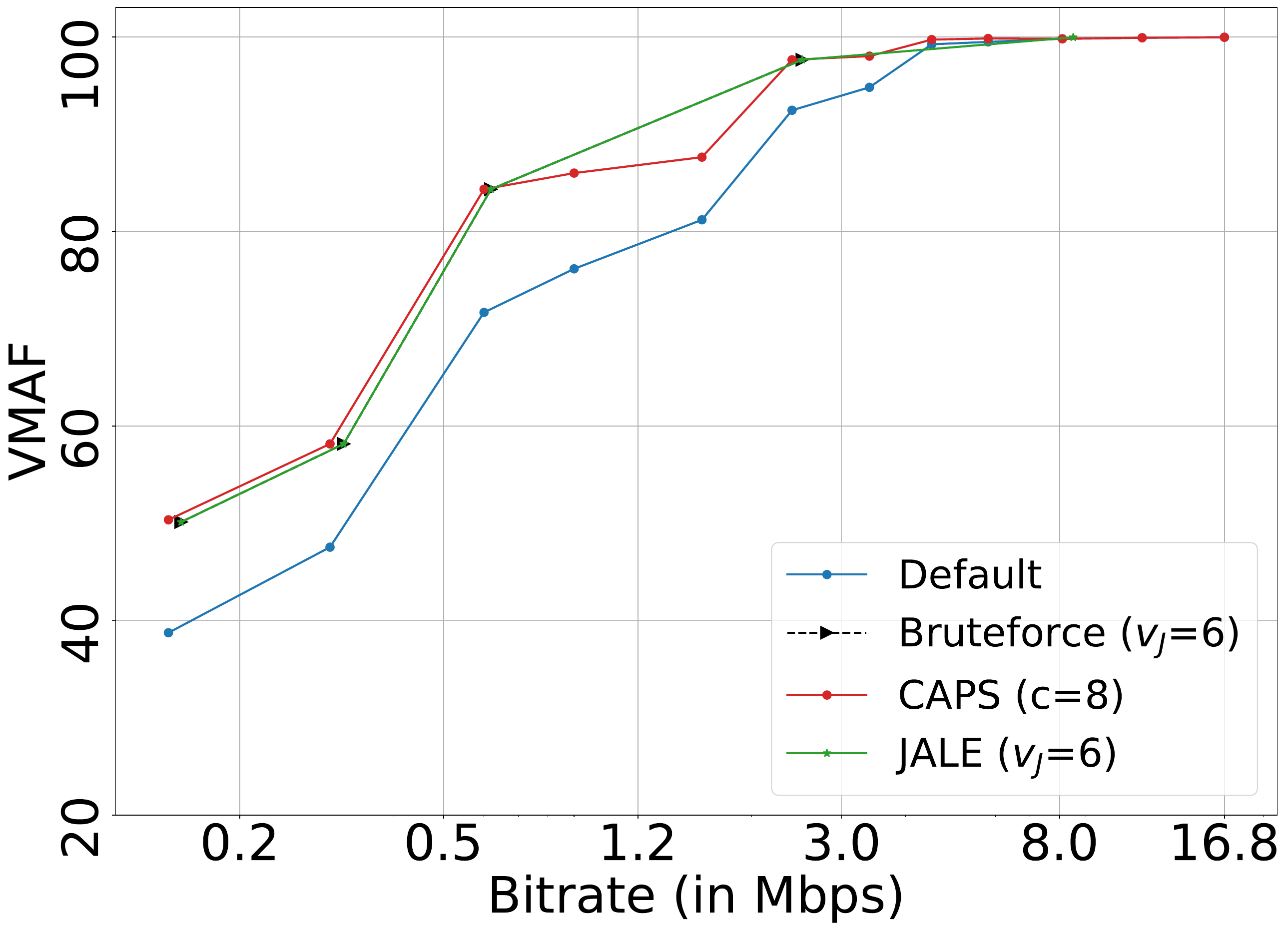}
    \includegraphics[width=\textwidth]{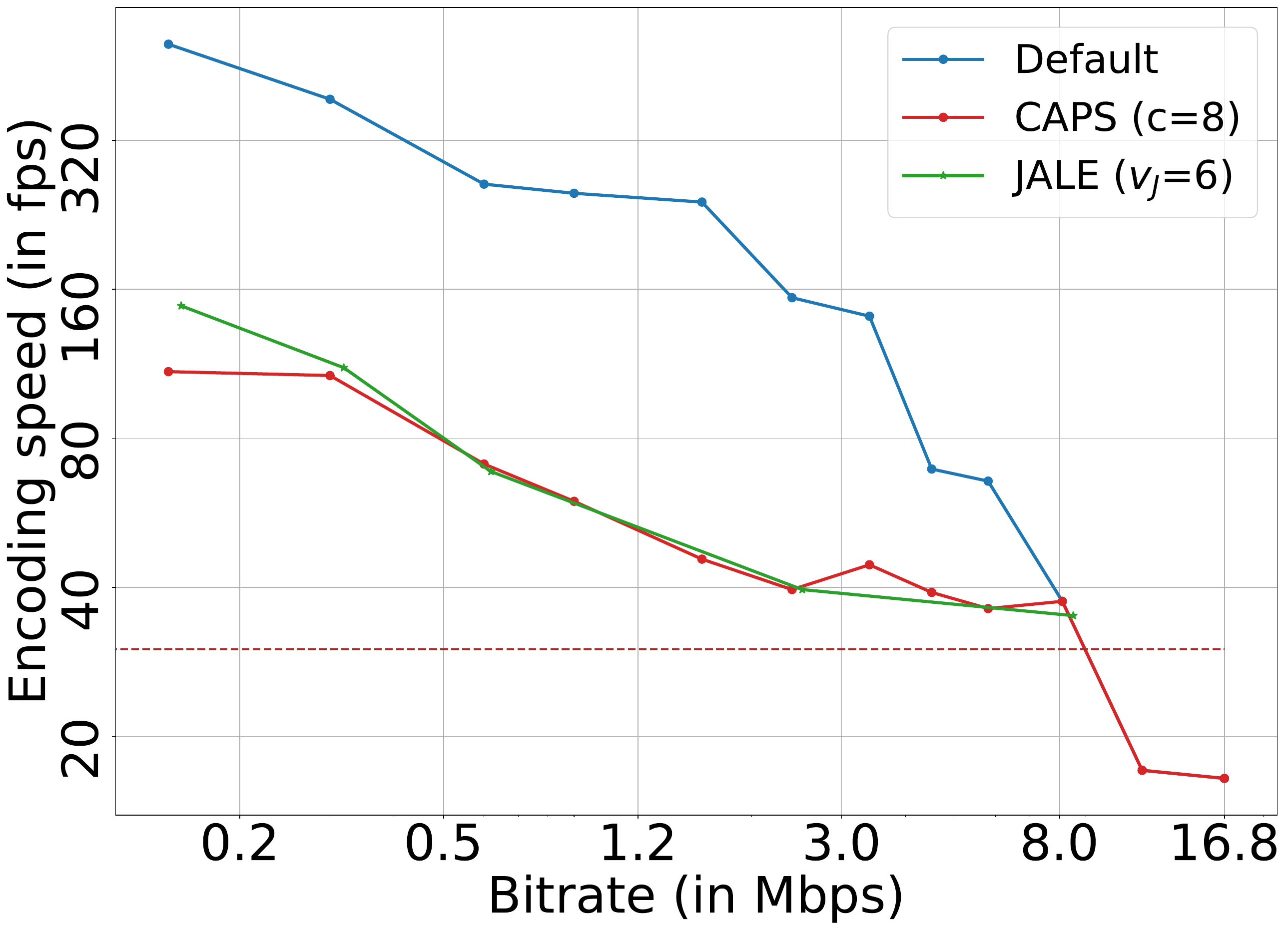}
    \caption{\textit{Characters\_s000}}
\end{subfigure}
\hfill
\begin{subfigure}{0.244\textwidth}
    \centering
    \includegraphics[width=\textwidth]{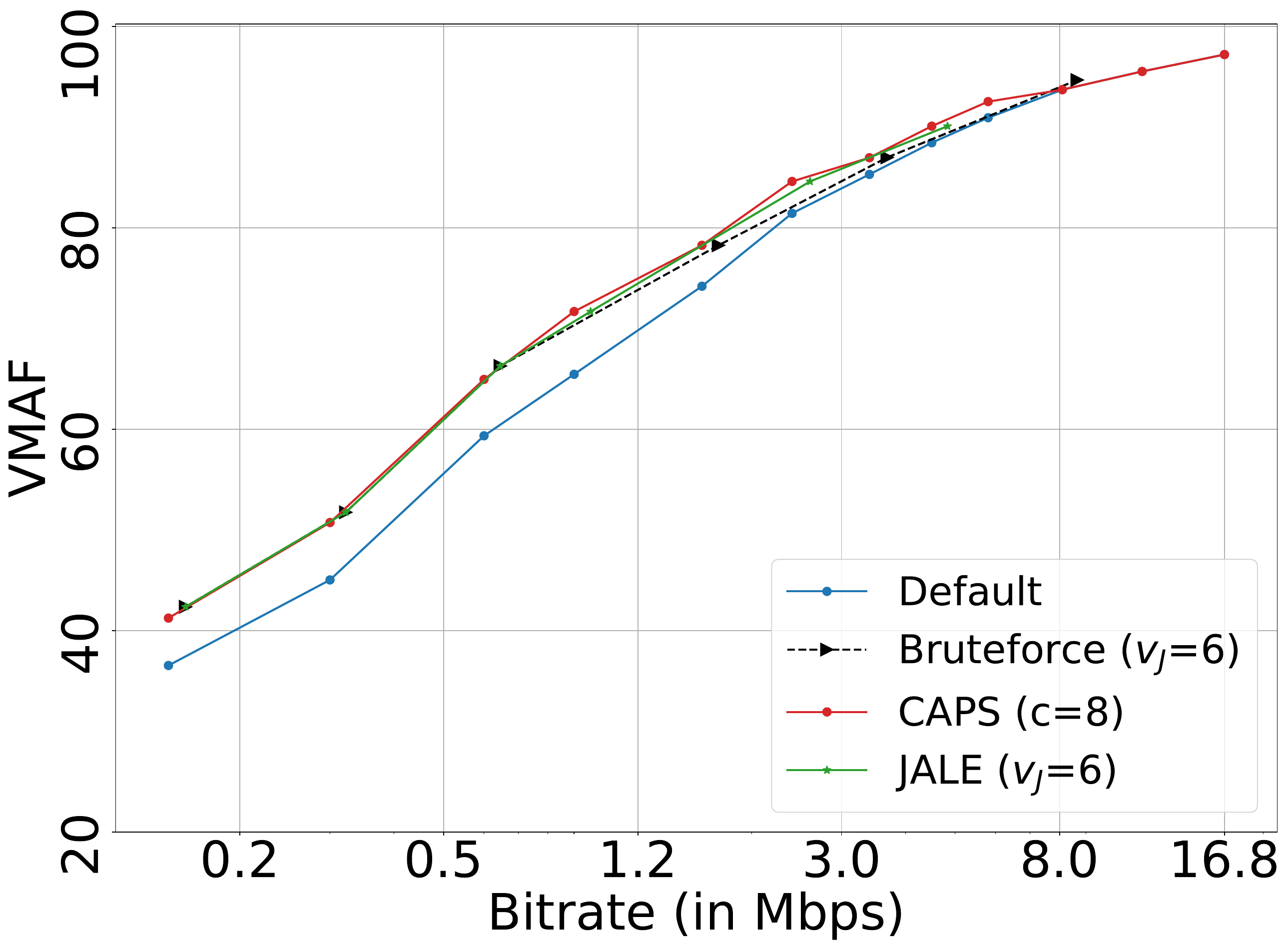}
    \includegraphics[width=\textwidth]{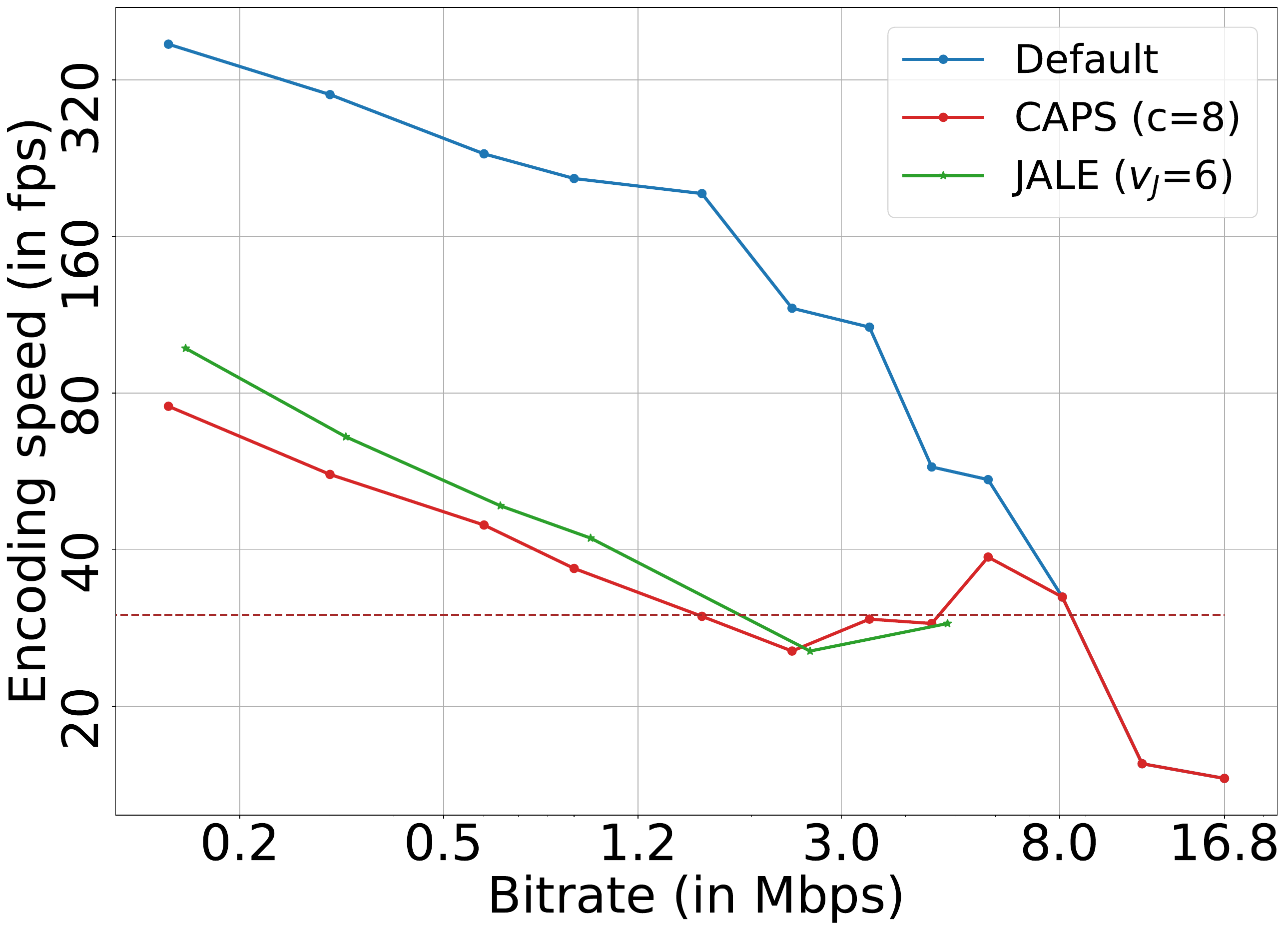}
    \caption{\textit{Dolls\_s001}}
\end{subfigure}
\hfill
\begin{subfigure}{0.244\textwidth}
    \centering
    \includegraphics[width=\textwidth]{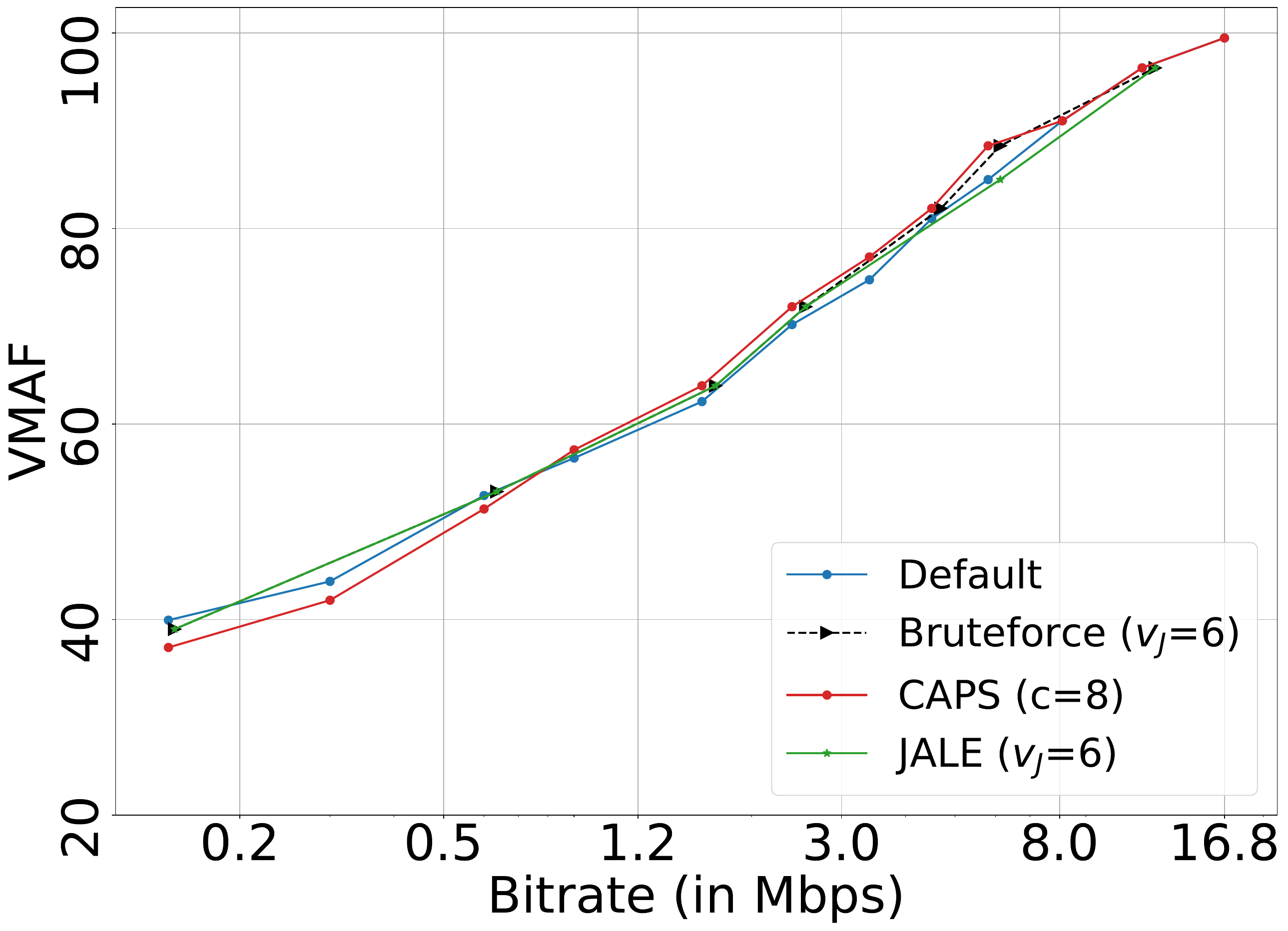}
    \includegraphics[width=\textwidth]{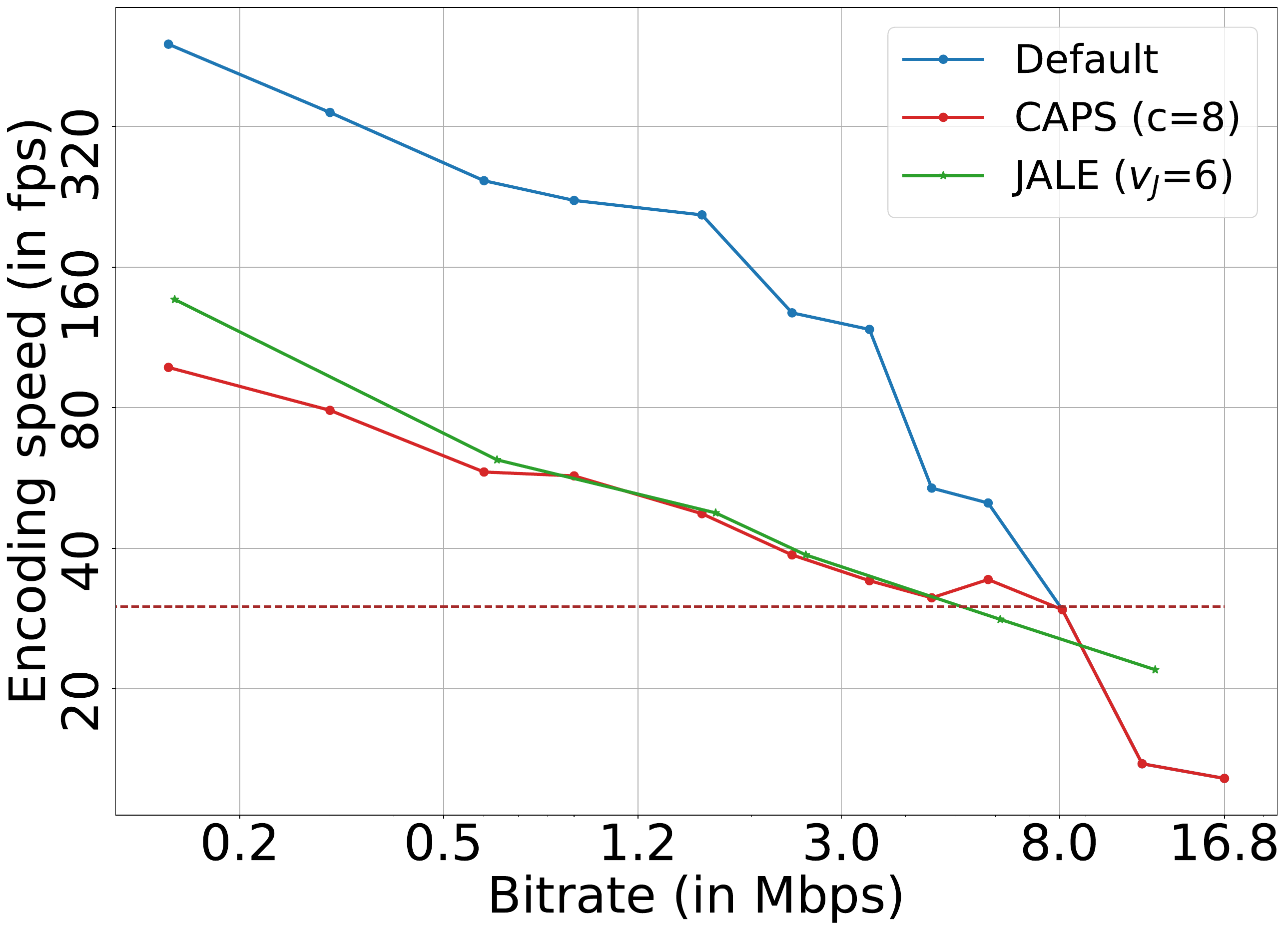}    
    \caption{\textit{Maples\_s000}}
\end{subfigure}
\vspace{-0.3em}
\caption{Rate-distortion (RD) curves of representative sequences (segments) for \emph{Default}~\cite{HLS_ladder_ref} encoding (blue line), \caps~($c=8$)~\cite{caps_ref} encoding (red line), compared to \ztps~(\vJ=6).}
\label{fig:rd_main_plot}
\end{figure*}

\textit{\textbf{Rate-distortion performance:}}
Figure~\ref{fig:vmaf_rep} shows the average VMAF for every representation in the bitrate ladder using the \emph{default}, \caps and \ztps encodings, while Figure~\ref{fig:rd_main_plot} shows the RD curves of the representative video segments of various video complexities with encoding using benchmark schemes, and \ztps~($v_{\text{J}}=6$). It is observed that the VMAF achieved by \ztps is higher than or close to \texttt{CAPS} encoding and consistently higher than the default encoding at the same target bitrates. Moreover, perceptually redundant representations are eliminated in \ztps. Furthermore, we observe a substantial improvement in quality at lower bitrate representations, owing to the selection of slower presets. 

\textit{\textbf{Bjøntegaard delta rates (BDR):}} 
We evaluate the coding efficiency using \BDRP, \BDRV, \mbox{BD-PSNR}, and \mbox{BD-VMAF} compared to the \emph{default} encoding, as shown in Table~\ref{tab:ztps_res_cons}. Bruteforce encoding~\cite{netflix_paper} yields \SI{100}{\percent} accurate results representing the highest bound of the compression efficiency improvement (in VMAF) compared to the \emph{default} encoding. Compared to the \emph{default} encoding, the coding efficiency improvement achieved by \ztps is similar to the bruteforce encoding. Hence, the prediction models used in \ztps~are deemed fairly accurate. 
Using \ztps~($v_{\text{J}}$=6), we observe an average bitrate reduction of \SI{25.93}{\percent}, \SI{25.47}{\percent} to maintain the same PSNR and VMAF. Furthermore, we observe an average quality improvement of \SI{1.32}{\decibel} PSNR and 5.38 VMAF points, respectively, at the same target bitrate.

\begin{table}[t]
\caption{Average results of the encoding schemes compared to the \emph{default} encoding.}
\centering
\resizebox{\columnwidth}{!}{
\begin{tabular}{l|c|c|c|c|c|c|c|c}
\specialrule{.12em}{.05em}{.05em}
\specialrule{.12em}{.05em}{.05em}
Method & \vJ & \BDRP & \BDRV & BD-PSNR & BD-VMAF & $\Delta S$ & $\Delta N$ & $\Delta E$\\
       &     &  [\%] & [\%]  & [dB]    &         &  [\%]      & [\%]       &  [\%]     \\
\specialrule{.12em}{.05em}{.05em}
\specialrule{.12em}{.05em}{.05em}
\multirow{3}{*}{Bruteforce} & 2  & -23.85 & -23.02 & 1.21 & 4.61 & -51.32 & -43.42 & 1688.91 \\
 & 4  & -24.54 & -24.27 & 1.25 & 5.26 & -62.13 & -53.08 & 1688.91 \\
 & 6  & -26.30 & -25.77 & 1.32 & 5.53 & -72.03   & -63.38 & 1688.91 \\
\specialrule{.12em}{.05em}{.05em}
\texttt{CAPS} ($c=4$) & - & -12.31 & -14.98 & 0.82 & 2.91 &  -0.06 & -50.00 & 8.65 \\
\texttt{CAPS} ($c=8$) & - & -20.26 & -20.95 & 1.06 & 4.03 &  -0.09 & 0 & 39.68 \\
\texttt{CAPS} ($c=16$) & - & -28.13 & -29.35 & 1.62 & 6.04 &  -0.35 & 100.00 & 96.84 \\
\specialrule{.12em}{.05em}{.05em}
\multirow{3}{*}{\ztps} & 2  & -23.30 & -22.80 & 1.19 & 4.47 & -51.34  & -43.13 & -2.69 \\
 & 4  & -24.41 & -24.08 & 1.25 & 5.05 & -62.95  & -54.08 & -22.70 \\
 & 6  & -25.93 & -25.47 & 1.32 & 5.38 & -72.70  & -63.83 & -37.87 \\
\specialrule{.12em}{.05em}{.05em} 
\specialrule{.12em}{.05em}{.05em}
\end{tabular}}
\label{tab:ztps_res_cons}
\end{table} 

\textit{\textbf{Storage consumption:}} 
We evaluate the relative difference in the storage space between the considered encoding schemes and the default encoding scheme to store all bitrate ladder representations. When the JND value increases, the number of representations in the bitrate ladder decreases, causing a decrease in the overall storage space needed for the representations of an input video segment. \ztps yields average storage reduction of up to \SI{72.70}{\percent} for various target \vJ values. 

\textit{\textbf{Encoding energy consumption:}} 
We conduct a comprehensive evaluation of encoding schemes by analyzing the relative differences in energy consumption during encoding ($\Delta E_{\text{enc}}$) of the bitrate ladder compared to the default encoding scheme. Predictably, compared to \emph{Default}, and \ztps encoding, \caps ($c=16$) yields the highest encoding energy consumption, owing to the selection of slower presets at all bitrates, as the constraint on total CPU threads is not considered. \ztps yields the lowest encoding energy compared to other benchmark methods. Moreover, as $v_{\text{J}}$ increases, the energy consumption is reduced.

\section{Conclusions}
\label{sec:conclusion_future_dir}
This paper proposed \ztps, a JND-aware low latency encoding scheme for adaptive live streaming applications. \ztps~jointly predicts the optimized encoder preset and CPU thread count for a given representation for each video segment based on the video content complexity features, target encoding speed, and the total number of available CPU threads. It helps improve quality and CPU utilization during encoding. Furthermore, the JND-based representation elimination algorithm removes perceptually redundant representations in the bitrate ladder. The performance of \ztps is analyzed using the x265 open-source HEVC encoder for the HLS bitrate ladder encoding. It is observed that \ztps yields an overall average quality improvement of \SI{0.98}{\decibel} PSNR and 4.41 VMAF points at the same bitrate, compared to \textit{ultrafast} encoding of the reference HLS bitrate ladder using eight CPU threads for each representation. Considering a JND of six VMAF points, storage, thread count, and encoding time reductions of \SI{72.70}{\percent}, \SI{63.83}{\percent}, and \SI{37.87}{\percent}, respectively, are observed. 

In the future, \ztps will support the addition of encoder presets beyond the pre-defined options, enhancing efficiency in the encoding process. This may ensure better flexibility in selecting encoding parameters such that the achieved encoding speed is the same as the target. Furthermore, storage reduction techniques and representation elimination to enhance the overall energy efficiency of video streaming and improve the sustainability of video streaming systems shall be investigated.

\section{Acknowledgment}
The financial support of the Austrian Federal Ministry for Digital and Economic Affairs, the National Foundation for Research, Technology and Development, Austrian Research Promotion Agency (FFG), grant agreement FO999897846 (GAIA) and the Christian Doppler Research Association is gratefully acknowledged. Christian Doppler Laboratory ATHENA: \url{https://athena.itec.aau.at/}.
\balance
\newpage
\balance
\bibliographystyle{IEEEtran}
\bibliography{references.bib}
\balance
\end{document}